# Design principles for optoelectronic light-scattering reservoir computing at the edge of chaos

*Geon Kim[1,2] and YongKeun Park[1,2,3,*]*
[1]Department of Physics, Korea Advanced Institute of Science and Technology (KAIST), Daejeon 34141, Republic of Korea
[2]KAIST Institute for Health Science and Technology, Korea Advanced Institute of Science and Technology (KAIST), Daejeon 34141, Republic of Korea
[3]Tomocube Inc., Daejeon 34051, Republic of Korea
[*]Corresponding author. E-mail: yk.park@kaist.ac.kr
ORCID: Geon Kim — [0000-0002-0661-8579]; YongKeun Park — [0000-0003-0528-6661]

Physical reservoir computing offers an energy-efficient route to sequential cognitive inference by outsourcing nonlinear temporal mixing to hardware substrates with rich intrinsic dynamics, with free-space light-scattering systems particularly attractive for their parallelism and reconfigurability — yet practical design principles linking hardware control variables to computational performance have remained unestablished. Here, we establish such principles by systematically mapping three physical control axes of a reconfigurable optoelectronic light-scattering reservoir — reservoir dynamics, input–reservoir coupling, and reservoir interconnectivity — and identifying a quantitative optimum along each axis. Within this design landscape, we observe a memory-capacity peak that coincides with near-zero maximal Lyapunov exponent and is quantitatively reproduced in numerical simulation, extending edge-of-chaos confirmations previously reported in ion-gating and spin-wave reservoirs into the photonic substrate. The two remaining axes exhibit a density–magnitude trade-off in input coupling and an intermediate optimum in reservoir interconnectivity. Operating at the resulting three-axis optimum, the reservoir achieves stable Mackey–Glass chaotic time-series prediction in free-running mode and 84.5% blind classification accuracy on the 10-class Speech Commands spoken-digit benchmark; the principles, stated in substrate-specific units yet rooted in substrate-independent concepts of criticality and balanced coupling, provide a transferable framework for reconfigurable optical reservoir hardware.

## 1. Introduction

Modern artificial intelligence is increasingly bottlenecked by the energy and latency cost of sequential inference. Current large-scale neural networks have achieved remarkable performance on a wide range of cognitive tasks by scaling models and datasets in tandem[1], but the computational and energetic demand of training and inference has grown commensurately, threatening the accessibility and sustainability of AI systems[2,3]. The challenge is particularly acute for sequential cognitive tasks — temporal forecasting[4], sequence recognition[5,6], and the processing of streaming data — for which standard digital architectures based on recurrent neural networks (RNNs) or attention mechanisms incur per-time-step computational overhead that scales unfavorably with sequence length[7]. In response, a parallel research effort has sought to outsource the repetitive sequential computation to physical substrates whose intrinsic dynamics naturally implement nonlinear temporal mixing, reading out the relevant features through a small trained interface. This motivation is aligned with the broader resurgence of deep optics and photonics for AI inference, where optical systems are being explored as fast and low-power domain-specific accelerators[8]. *Physical reservoir computing* — the broad class of such schemes — has demonstrated promising performance and energy efficiency across diverse substrates, including ion-gating transistors[9], spin-wave systems[10], atomic-switch nanowire networks[11], fiber optics feedback loops[12,13], silicon-photonic integrated circuits[14,15], and free-space optical scattering[16–18].

The principle underlying physical reservoir computing is that a dynamical system with fading memory, driven by an external input, generates a high-dimensional internal state from which a wide variety of temporal features can be extracted by a simple readout[19,20]. Unlike a conventional RNN, which is a precursor of reservoir computing where all weight matrices are learned via backpropagation through time, a reservoir computing system holds the recurrent and input weights fixed and trains only the readout. This dramatically simplifies the optimization and aligns naturally with hardware substrates whose recurrent dynamics are physically determined rather than numerically learned. This advantage, however, comes with an open methodological question: in the absence of training, the computational performance of a physical reservoir depends sensitively on the choice of the substrate, which can vary in dynamical regime, the input-reservoir coupling, and the connectivity of the recurrent operator[7,21]. Theoretical and experimental studies of cellular automata[22] and RNN-based reservoirs[23,24] have long suggested that an optimal operating regime exists at the *edge of chaos* — the boundary between dynamical stability and chaos — but the experimental confirmation of this prediction in physical reservoirs is recent and substrate-specific[9,10]. For optical reservoir computing, in particular, tuning of control parameters have been reported[12–14], but with limited tuning options that were not linked directly to reservoir computing principles.

Optical reservoir computing offers an attractive substrate for the design-landscape mapping we develop here. Free-space scattering reservoirs in particular are scalable to thousands of nodes through the use of spatial light modulators and image sensors[16–18], and are reconfigurable in terms of recurrence intensity, input injection, and recurrence structure which is difficult to achieve in fiber optics[12] or silicon photonics[14] implementations. This use of scattering differs from transmission-matrix-based wavefront shaping, where the complex optical response of a scattering medium is explicitly calibrated to enable deterministic focusing, image transmission, or inversion through the medium[25–27]. Here, the scattering layer is not measured, inverted, or trained; it is used as an untrained recurrent mixing operator whose computational utility is determined by reservoir-level dynamical observables. This reconfigurability is the key practical asset for design-landscape mapping. We exploit this asset to perform a systematic three-axis exploration of an optoelectronic light-scattering reservoir. The reconfigurable scattering reservoir should not be conflated with deep diffractive neural networks[28,29], whose scattering layers are numerically *trained* to perform a specific task, nor genetic-algorithm-optimized optical RNNs[30], in which the reservoir or the readout is task-tailored. The present work concerns the *untrained* light-scattering reservoir, whose practical design has thus far been guided largely by intuition and parameter searches that are specific to the hardware[18].

In this work, we establish design principles for light-scattering reservoir computing by systematically mapping three physical control axes — reservoir dynamics, input–reservoir coupling, and reservoir interconnectivity. Within this design landscape, we identify quantitative optima along each axis, including a memory-capacity peak that coincides with near-zero maximal Lyapunov exponent — extending edge-of-chaos confirmations previously reported in ion-gating and spin-wave reservoirs into the photonic substrate. We further validate the resulting three-axis design optimum on two sequential cognition tasks — Mackey–Glass chaotic time-series prediction in free-running mode and 10-class Speech Commands spoken-digit recognition with 84.5% blind classification accuracy — and discuss the prospects for extending these principles to other reconfigurable optical reservoir hardware, including fully-optical cavity-based and integrated-photonic platforms.

## 2. Results

### *2.1. Light-Scattering Reservoir Computing Platform*

We implement reservoir computing in a free-space optoelectronic system whose three control axes — detector exposure time, the input-reservoir weighting, and the scatterer phase profile — span the design landscape that we characterize in the remainder of this section. We describe reservoir operations (Figure 1a) with the following equations.

$$\mathbf{h}(t) = g\big(\mathbf{W}_{\text{rec}} f(\mathbf{h}(t-1), \mathbf{x}(t); \mathbf{W}_{\text{in}})\big) \quad \text{(Eq. 1)}$$

$$\mathbf{y}(t) = \mathbf{W}_{\text{out}} \mathbf{h}(t) \quad \text{(Eq. 2)}$$

Equation (1) illustrates the recurrent update of the reservoir state vector $\mathbf{h} \in \mathbb{R}^{M}$, given the input vector $\mathbf{x} \in \mathbb{R}^{L}$. $f$ denotes the input-reservoir coupling operation, parametrized by $\mathbf{W}_{\text{in}} \in \mathbb{R}^{M\times L}$. $\mathbf{W}_{\text{rec}} \in \mathbb{R}^{M\times M}$ is the matrix that interconnects between elements of $\mathbf{h}$ or the reservoir nodes. $g$ is the nonlinear operation that yields the updated reservoir state. Equation (2) is the readout of the output vector $\mathbf{y} \in \mathbb{R}^{N}$ from the reservoir state $\mathbf{h}$. Note that the readout matrix $\mathbf{W}_{\text{out}} \in \mathbb{R}^{N\times M}$ is the only component of the reservoir operation that is numerically trained.

Our system maps the recurrent computation of Equation (1) to a sequence of physical operations. The reservoir state and input are coupled and projected on the display as the phase of a coherent optical field $\mathbf{E}$ and subsequently scattered into $\mathbf{E}_{\mathrm{scat}}$ by the configurable scatterer — realized by another SLM at the aperture plane — to be incident on the CMOS camera, which records the intensity profile (Figure 1b; see supplementary Figure S1a for the technical illustration with all optical components). This process can be written as

$$\mathbf{h}(t) = \min\left(\left|\mathbf{W}_{\mathrm{rec}} \exp\left(i\pi\left(\left(\mathbf{I} - \widetilde{\mathbf{W}}_{\mathrm{in}}\right)\mathbf{h}(t-1) + \widetilde{\mathbf{W}}_{\mathrm{in}}\tilde{\mathbf{x}}(t)\right)\right)\right|^2, 1\right) \quad \text{(Eq. 3)}$$

where $f$ is replaced with weighted summation in phase representation and $g$ is replaced with intensity measurement (see supplementary Figure S1b for visual description), from Equation (1). We characterize the system as an optoelectronic reservoir as the phase-encoding of $\mathbf{h}(t-1)$ and $\mathbf{x}(t)$ in $f$, as well as the intensity recording in $g$ is performed *electronically*. The relation to fully-optical implementations is taken up in Section 3.3. The three physical parameters that we vary throughout the study — detector exposure time (which sets the norm of $\mathbf{W}_{\mathrm{rec}}$), distribution of nonzero $\mathbf{W}_{\mathrm{in}}$ entries, and scatterer diffusing angle — map directly to the design axes shown schematically in Figure 1c. Each parameter can be swept independently of the others by changing only displayed phase patterns or camera settings, with no modification of the optical hardware.
In line with the standard reservoir computing protocol[19,21], the readout weight matrix $\mathbf{W}_{\mathrm{out}}$ in Equation (2) is obtained by taking the Moore–Penrose pseudoinverse between the ground truth outputs and reservoir states across the training time points. All training is therefore confined to a single linear mapping. Further detail on the numerical training process of $\mathbf{W}_{\mathrm{out}}$ is described in Methods §4.7.
We also arrange the input coupling process to be able to accommodate various cases of input dimension $L$. In specific, each reservoir node is either irrelevant with input coupling or assigned a single element of $\mathbf{x}(t)$, whose index cycles throughout the nodes that are coupled with the input. This is equivalent to setting a diagonal matrix $\widetilde{\mathbf{W}}_{\mathrm{in}} \in \mathbb{R}^{M\times M}$ that satisfies

$$\mathbf{W}_{\mathrm{in}}\mathbf{x}(t) = \widetilde{\mathbf{W}}_{\mathrm{in}}\tilde{\mathbf{x}}(t) \quad \text{(Eq. 4)}$$

where each diagonal entry of $\widetilde{\mathbf{W}}_{\mathrm{in}}$ corresponds to a reservoir node, and $\tilde{\mathbf{x}}(t)$ is repetition of $\mathbf{x}(t)$ elements throughout the indices of nonzero $\mathrm{diag}\left(\widetilde{\mathbf{W}}_{\mathrm{in}}\right)$ elements (see Equation (8) and (9) of Methods 4.4 for detailed description of $\tilde{\mathbf{x}}(t)$ and $\widetilde{\mathbf{W}}_{\mathrm{in}}$). Note that $\widetilde{\mathbf{W}}_{\mathrm{in}}$ is reshaped into $\mathbf{W}_{\mathrm{in}}$ for applications involving different dimensions of input sequence. Therefore, the indices and values of nonzero $\mathrm{diag}\left(\widetilde{\mathbf{W}}_{\mathrm{in}}\right)$ elements determine which nodes are coupled with input and how dominantly the input is coupled over the existing state, respectively. Throughout the study, we set one single value within $[0,1]$ across all nonzero diagonal elements to achieve even coupling in all input dimensions and reservoir node indices. We adopt phase modulation across both the input encoding and the reservoir state to represent a continuous-valued reservoir state, maintain reservoir activations within a physically controllable range, and effectively utilize the available dynamic range; binary intensity modulation via digital-micromirror devices is discussed as a higher-throughput alternative in Section 3.5.

*2.2. Reservoir Dynamics and Criticality*

A central question in physical reservoir computing is the existence and identification of an operating regime in which the reservoir best preserves the past inputs while remaining responsive to new ones. One of the earliest studies defines an identifying property of reservoir computing systems as the uniqueness of $\mathbf{h}(t)$ for a left-infinite input history $\{\mathbf{x}(t-t')\}_{t'=0}^{\infty}$[19]. In terms of Lyapunov stability (Methods 4.5), this is closely related to a non-positive autonomous (zero-input) maximal Lyapunov exponent $\lambda_{\max} \leq 0$. Whereas the echo-state property formally concerns convergence under input drive, the autonomous $\lambda_{\max}$ serves as an experimentally accessible measure for the reservoir‘s dynamical property. On the other hand, different studies have reported that this effective regime is highly associated with the boundary between dynamical stability and chaos[22–24] — known as the *edge of chaos* — characterized by $\lambda_{\max}$ near 0. Recent experiments confirm this hypothesis in two physical substrates. Ion-gating reservoirs based on a lithium-ion electrolyte/diamond interface exhibit dynamics with $\lambda_{\max} \approx 0$ at specific transistor channel lengths[9]. Spin-wave interference-based reservoirs exhibit a waveform-transformation peak coinciding with $\lambda_{\max} \approx 0$ as the pulse interval is tuned[10].
Here, we map this regime of critical dynamics in the optical domain. To do so, we identify the camera exposure time $T$ as the most direct experimental control over reservoir dynamics: as Equation (3) indicates, the rate at which a perturbation in $\mathbf{h}(t)$ evolves over time is set primarily by the norm of $\mathbf{W}_{\text{res}}$, controlled experimentally by $T$ (Methods 4.2, 4.5).
Sweeping $T$ over an order of magnitude in a 400-node (20×20 pixel) reservoir accompanies a traversal of dynamical regimes. $\lambda_{\max}$ measured by the perturbation-propagation protocol of Methods 4.5, increases from negative values ($\lambda_{\max} < 0$, perturbations to a reservoir state decay) through criticality ($\lambda_{\max} \approx 0$) into the chaotic regime ($\lambda_{\max} > 0$, perturbations diverge exponentially). Note that the increase of $\lambda_{\max}$ may not be monotonic, as increased exposure may lead to pixel saturation at the camera which stabilizes the reservoir dynamics. Numerical simulation (Figure 2a, black lines) and experiment (Figure 2a, red lines) display similar changes of $\lambda_{\max}$ across the explored range of $T$. The dynamical transition is also visible in the reservoir state changing over time under zero input: at $\lambda_{\max} < 0$ the trajectory collapses toward a significantly narrow range or a fixed point; at $\lambda_{\max} \approx 0$ it traces a larger yet structured variation; at $\lambda_{\max} > 0$ it diverges with little correlation to the previous time points (Figure 2d). These three regimes match the canonical predictions of low-dimensional dynamical systems theory[24] and provide the dynamical substrate on which we now examine memory performance.
The memory capacity (MC), defined as the cumulative reconstruction fidelity of past input history (Equation (12), Methods 4.6), changes sharply with $T$, MC for both simulation and experiment forms a unimodal curve with maximum at a consistent range of $T$ value that achieves $\lambda_{\max} \approx 0$ (Figure 2b). When MC is plotted directly against $\lambda_{\max}$ the simulation and experimental data further converge to a sharp peak at $\lambda_{\max} \approx 0$ (Figure 2c), replicating the high computational performance at criticality reported in previous works[24]. The simultaneous coincidence of the peak between simulation and experiment is consequential. It demonstrates that (i) the optical reservoir dynamics aligns with the numerical modeling, providing interpretability and supporting efficient exploration of regimes in small scales; and (ii) the location of the MC peak is not an artifact of finite measurement noise or hardware imperfection but a property of the underlying dynamics.

To our knowledge, no prior optical reservoir-computing platform — whether based on fiber-loop[12], silicon-photonic[14,15], or free-space scattering[16–18,31] architectures — has reported a direct, controlled measurement of $\lambda_{\max}$ accompanying an MC peak. Our result therefore positions photonics as the third experimental substrate at which the edge-of-chaos hypothesis is confirmed in physical reservoir computing, joining ion-gating[9] and spin-wave[10] reservoirs. Beyond providing a substrate-independent design rule, this universality has a practical consequence for the design landscape of optical reservoir computing. It identifies the dynamics axis as the dominant control variable and indicates that the optima for the remaining axes — input coupling and reservoir interconnectivity — should be characterized at $\lambda_{\max} \approx 0$ throughout the rest of this work.

### *2.3. Input–Reservoir Coupling*

Having identified reservoir dynamics as the dominant control axis and setting the exposure time $T$ at the value yielding $\lambda_{\max} \approx 0$, we next examine the role of input coupling. Whereas reservoir dynamics determines whether information persists in the reservoir, input coupling determines how external information is injected into it: the same dynamical regime can result in different MC depending on two input-related parameters — density and magnitude of $\mathbf{W}_{\text{in}}$. We therefore sweep these two parameters while maintaining $\lambda_{\max} \approx 0$ in 400-node reservoirs as in Section 2.2, and ask whether the input-related design landscape exhibits a similarly well-defined optimum.

As Equation (3) and (4) make explicit, the input weight matrix $\mathbf{W}_{\text{in}}$ serves a dual role: it specifies which reservoir nodes receive the external input, and at what magnitude. We accordingly parameterize $\widetilde{\mathbf{W}}_{\text{in}}$ by two scalar quantities — the nonzero diagonal entry ratio $\rho$ (the fraction of reservoir nodes coupled with input) and the nonzero entry weight $w$ — and treat them as the two scalar control parameters of the input-reservoir coupling domain. To be more precise, one can express $\rho$ and $w$ as the following.

$$\rho = \left\|\text{diag}\left(\widetilde{\mathbf{W}}_{\text{in}}\right)\right\|_0 / M \quad \text{(Eq. 5)}$$

$$\text{diag}\left(\widetilde{\mathbf{W}}_{\text{in}}\right)_{\boldsymbol{i}} \in \{0, w\} \quad \text{(Eq. 6)}$$

To minimize local variability arising from specific node-selection patterns, the nonzero positions are uniformly distributed across the reservoir's pixel coverage; the elements of input $\mathbf{x}$ are rescaled to $[0,1]$ as described in Methods 4.4 (Figure 3a). The two-parameter sweep is performed in the same 400-node reservoir as Section 2.2, with exposure time at the value that yields $\lambda_{\max} \approx 0$.

The resulting MC landscape exhibits a high-performance region in the coupling parameter plane (Figure 3b). MC is suppressed at both low $\rho$ and $w$, where the recurrent dynamics dominate and external information is insufficiently registered in the reservoir state. With $\rho$ and $w$ both high, MC is also suppressed, but for the opposite reason: the input now drives nearly all reservoir nodes resulting in the silencing of earlier input history. The high-MC region traces an intermediate band along which the input-reservoir coupling is balanced, among which we identify the optimal operating point ($\rho = 21\%$, $w = 0.25$; green star in Figure 3b). Plotting MC along the profile of $\rho$ and $w$ crossing the operating point displays qualitative agreement between the experiment and simulation (Figure 3c). Beside the agreement, the experimental high-MC region is shifted slightly toward larger $\rho$ and smaller $w$, consistent with the residual hardware imperfections (super-pixel matching error, SLM phase quantization, camera read noise, scatterer alignment) that effectively reduce the per-pixel modulation depth. Having fixed both the dynamical regime and the input-coupling optimum, we next ask whether a similar quantitative design rule emerges along the remaining two axes — reservoir scale and scatterer interconnectivity.

*2.4. Reservoir Interconnectivity*

The remaining axis of reservoir interconnectivity controls how the reservoir state is intermixed across nodes during each recurrent operation, and therefore is governed by the structure of $\mathbf{W}_{\text{rec}}$ in Equation (1). Throughout this section, the reservoir dynamics is held at $\lambda_{\max} \approx 0$ and the input coupling at the optimum $(\rho, w) = (21\%, 0.25)$ identified in Section 2.3.

Prior to sweeping the interconnectivity through altering the scatterer profile, we increased the size of our reservoir $M$ to 10,000 which corresponds to the number of super-pixels contained by the largest square region in the display, i.e. $100 \times 100$ (Figure 4a). This is to study reservoir interconnectivity at the largest available scale, since the light scattering reservoir interconnectivity is influenced by both the scale and scatterer profile; interconnectivity of small-scale reservoirs grows faster than that of larger-scale reservoirs while increasing the scatterer diffusing angle (Figure 4b). MC increases concavely with reservoir scale across the swept range (Figure 4a). $M$ is varied from 100 ($10 \times 10$ pixels) to 10,000 ($100 \times 100$ pixels), corresponding to a hundredfold increase; while MC grows steeply for small $M$, the increase is more moderate above approximately $M = 3{,}600$ ($60 \times 60$ pixels), in both simulation and experiment. We adopt the upper end of the swept range, $M = 100 \times 100 = 10{,}000$ nodes, to ensure the highest potential performance in the following cognitive-task demonstrations of Section 2.5.

Scatterer interconnectivity is controlled experimentally by the diffusing angle $\theta$ of the SLM scatterer, generated by the iterative phase-profile algorithm of Methods 4.3. MC at $\lambda_{\max} \approx 0$ is non-monotonic in $\theta$ and exhibits a moderate maximum at $\theta = 0.2°$ for both the simulation and experiment (Figure 4c). Sweeping $\theta$ from 0.1° to 0.7° spans a range over which the density — especially off-diagonal in our demonstration — of the effective recurrent matrix $\mathbf{W}_{\text{rec}}$ varies from low (each input super-pixel illuminates roughly itself plus immediate neighbors) to high (each super-pixel illuminates an extended speckle pattern across the camera) (Figure 4d).

This mild optimal range of reservoir interconnectivity, together with the edge-of-chaos peak of Section 2.2 and the trade-off optimum of Section 2.3, constitutes the three quantitative design guidelines. Having identified each design optimum and characterized its quantitative shape, we now ask the practical question: does operating at this three-axis optimum deliver meaningful performance on cognitive tasks?

*2.5. Chaotic Time Series Prediction*

To validate the design principles in a synthetic dynamical setting, we deploy the optical reservoir at the three-axis optimum — $\lambda_{\max} \approx 0,\ (\rho, w) = (21\%, 0.25),\ \theta = 0.2°$ — on Mackey-Glass time-series prediction. We generate the time series by integrating the time-delayed differential equation

$$\frac{\mathrm{d}x(t)}{\mathrm{d}t} = \frac{ax(t-\tau)}{1 + x(t-\tau)^n} - b\,x(t) \quad \text{(Eq. 7)}$$

with $a = 0.2,\ b = 0.1,\ n = 10$, and $\tau = 17$, the canonical chaotic regime in which two trajectories with arbitrarily close initial conditions diverge over time[4]. The reservoir is trained to predict $x(t+1)$ from its current state, and prediction quality is assessed in two modes: a next-step mode in which the ground-truth $x(t)$ is fed to the reservoir at every step, and a free-running mode in which the reservoir's own one-step prediction is fed back as the next-step input. Free-running prediction is the more demanding of the two: errors compound, and the reservoir must reproduce the chaotic attractor as a closed-loop dynamical system rather than as a feedforward predictor.

The three dynamical regimes of Section 2.2 produce qualitatively different prediction outcomes (Figure 5). In the chaotic regime ($\lambda_{\max} > 0$, Figure 5a), both modes are inaccurate, with mean squared errors of $7.28 \times 10^{-3}$ next-step and $1.84 \times 10^{-1}$ free-running, and the free-running phase-space trajectory of $(x(t), x(t-\tau))$ is jagged and visibly discontinuous between consecutive steps. In the stable regime ($\lambda_{\max} < 0$, Figure 5b), the next-step mode is relatively accurate (mean squared errors of $1.23 \times 10^{-4}$) but the free-running mode still displays a phase-space behavior deviating from the actual time series (mean squared errors of $1.12 \times 10^{-1}$). Only at the edge of chaos ($\lambda_{\max} \approx 0$, Figure 5c) do both modes show significant resemblance to the time series: the next-step error is small ($1.18 \times 10^{-4}$) and the free-running trajectory remains bounded over the test horizon of 250 time steps ($3.44 \times 10^{-2}$), with the phase-space trajectory occupying a similar region with the ground-truth attractor. The chaotic Mackey-Glass dynamics is therefore reproducible by the optical reservoir only when the reservoir itself is dynamically critical — a finding consistent with the MC peak of Section 2.2. With the validation of temporal prediction in place, we now turn to a real-world cognitive benchmark.

*2.6. Spoken Digit Recognition*

The spoken-digit recognition task demonstrates that the same critical-dynamics design optimum supports practical sequential classification. Using the high-performance design shown in Section 2.5 (Figure 5c), we tackle the 10-digit subset of the Speech Commands dataset[6], comprising 200 distinct speakers per digit; for each digit we randomly hold out 20 utterances as a blind test set, training the readout on the remaining 180. Each utterance is preprocessed into a 32-channel log-mel spectrogram (Methods 4.4) and streamed sequentially through the input encoding of Figure 3 to the reservoir (Figure 6a). The reservoir state at each train time step is collected and a linear readout is trained by Moore-Penrose pseudoinverse (Methods 4.7) to map each $\mathbf{h}(t)$ to the corresponding output $\mathbf{h}(t)$, of which the ground truth is a one-hot class label that represents the digit being streamed at $t$. At test time steps, the predicted class is obtained by averaging the readout output over the duration of each utterance and selecting the maximal-response class (Figure 6b).

The optical reservoir achieves an overall blind classification accuracy of 84.5% across the 10 digits, with a confusion matrix that is strongly diagonal and concentrates its residual off-diagonal mass on a small number of acoustically confusable digit pairs (Figure 6c). The most frequently misrecognized utterances are "two" classified as "three", followed by "eight" classified as "three" or "six"; the digits "zero", "two", and "eight" produce no false-positive predictions in the test set. The most reliably recognized digit is "three" (100% recall), and the least reliable is "eight" (60% recall).

The resulting performance shows that operating at the three-axis design optimum is sufficient to reach the regime of practical sequential classification on an external natural world database. This highlights the design principles to facilitate light scattering reservoirs to handle real-world applications rather than specifically tackling numerical benchmark tasks.

Together, the chaotic time-series prediction and spoken-digit results validate the design principles established in Sections 2.2 through 2.4 in two complementary settings — reproducing numerically modeled chaotic motion and discrete categorical decision — and motivate a closer examination of what these principles imply for the broader landscape of optical reservoir computing.

## 3. Discussion

### *3.1. Three Design Principles for Light-Scattering Reservoir Computing*

The systematic mapping of Sections 2.2–2.4 reveals three quantitative design principles that govern the computational performance of light-scattering reservoir computing. First, the reservoir must be operated at the edge of chaos, with the maximal Lyapunov exponent $\lambda_{\mathrm{max}}$ tuned to approximately zero by the overall light intensity reaching the camera; deviations of $\lambda_{\mathrm{max}}$ in either direction degrade memory capacity sharply. Second, the input weight matrix $\mathbf{W}_{\mathrm{in}}$ should occupy a domain of trade-off in the (density $\rho$, magnitude $w$) plane where input driving and recurrent dynamics maintain balance, with the optimum for our phase-encoding regime sitting at a region spanning around $\rho = 21\%$ and $w = 0.25$. Third, scatterer interconnectivity — controlled here by the diffusing angle of the scatterer's phase profile — should be at an intermediate level ($\theta = 0.2°$ for our reservoir of $M = 10{,}000$) which leads to information propagating between reservoir nodes at a moderate breadth. These three principles are stated in substrate-specific units ($T, \ \rho, \ w, \ \theta$), but their underlying principles — peak at criticality, balanced input coupling, and intermediate connectivity — should transfer to other reconfigurable optical reservoir architectures, including amplitude-modulated free-space systems and integrated photonic platforms in which the reservoir state is encoded on coupled waveguides or microring resonators.

### *3.2. Edge of Chaos as a Universal Organizing Principle of Physical Reservoir Computing*

The most decisive of the three design principles was the critical stability of $\lambda_{\mathrm{max}} \approx 0$, which displayed a sharp peak of MC. While this MC peak at criticality had been predicted in theoretical and simulated studies[22–24] its experimental confirmation in physical reservoir computing has accumulated only recently and substrate by substrate. Nishioka et al.[9] reported $\lambda_{\text{max}} \approx 0$ specific to the transistor channel length, in an ion-gating reservoirs based on an electric double layer mechanism. Nishimura et al.[10] subsequently observed a similar peak in a spin-wave reservoir realized on yttrium iron garnet, with the dynamics tuned by input pulse interval. Our observation in a free-space light-scattering reservoir, with the dynamics tuned by detector exposure time, joins ion-gating and spin-wave reports as a third experimental confirmation in a physically distinct substrate. Together, these observations support the edge of chaos as a useful organizing principle for memory-dominated operation in physical reservoir computing; whether the principle generalizes to all task-specific performance regimes across substrates remains an open question to be characterized per substrate. The practical implication for designers of new physical reservoirs is direct: the dynamics axis should be characterized first, by measuring or estimating $\lambda_{\text{max}}$ as a function of the most accessible physical control variable, and the remaining design axes should be optimized only at the resulting $\lambda_{\text{max}} \approx 0$ operating point. In our system the dynamics axis emerged as the dominant control variable: the MC range covered by sweeping the camera exposure was substantially larger than that covered by sweeping either of the remaining two axes (Figures 2–4). We expect this dominance to be a general feature of physical reservoir computing, although confirmation in additional substrates remains an open question.

*3.3. Scope: Optoelectronic vs. Fully-Optical Reservoir Computing*

We note explicitly that the present system is, strictly speaking, an *optoelectronic* rather than a fully-optical reservoir computer: as made explicit by the three-component decomposition of Equation (3), the reservoir state at each time step is detected by a camera, digitized, and re-encoded by the phase-encoding onto the display SLM introducing an electronic round-trip and the associated latency. The scatterer represented by $\mathbf{W}_{\text{in}}$ supplies the light-speed mixing between reservoir nodes within each time step, but the inter-step recurrence is closed by the camera-to-SLM feedback path rather than by a passive optical cavity. Fully-optical scattering reservoirs in which the modulator and the scattering medium are placed inside an optical cavity, so that the reservoir state evolves continuously in the optical domain without electronic intermediation, more closely embody the limit of pure-optical reservoir computing and have been pursued in cavity-based architectures by Sunada and co-workers[32]. A closely related line of work by Davidson, Friesem, and co-workers at the Weizmann Institute demonstrates intracavity-SLM control of multi-mode dynamics in degenerate cavity lasers, where the same modulator-plus-disorder ingredient is embedded inside the optical cavity[33,34]. Their SLM-tuned cavity disorder plays a role analogous to our reconfigurable scatterer, supporting the expectation that the three design principles transfer to cavity-internal, fully-optical reservoir-computing platforms. The objective of the present work is not, however, to maximize bandwidth or to minimize latency — the parameters along which fully-optical implementations would clearly outperform our optoelectronic prototype — but to characterize the design landscape that governs the computational performance of light-scattering reservoirs. For this objective, the optoelectronic configuration provides direct experimental access to all three design axes: detector exposure tunes the dynamics, the input SLM weighted mapping tunes input-reservoir coupling, and the scatterer angular profile tunes interconnectivity. The same three design axes exist in fully-optical implementations — they are simply controlled by different physical handles (cavity gain, intra-cavity modulator, and intermode coupling, respectively) — and we expect the three design principles articulated in Section 3.1 to inherit.

*3.4. Position with Respect to Recent Photonic Reservoir Computing*

The recent optical computing literature has made impressive progress along several complementary axes. At the broader level of optical AI inference, deep optics and photonics have been articulated as a route toward fast, low-power, domain-specific inference engines[8], and multimode-fiber-based optical learning operators have shown that complex spatiotemporal mode interactions can serve as scalable physical computation engines for classification and regression tasks[35]. Within photonic reservoir computing more specifically, progress has proceeded along two complementary axes that are largely orthogonal to the present work. On the throughput axis, Wang and co-workers demonstrated a silicon-photonic reservoir computer reaching 200 TOPS[15], pushing optical reservoirs into the regime where the per-operation energy cost approaches the limits set by photonic-electronic conversion. On the algorithmic axis, the group that first implemented free-space light-scattering reservoir computing subsequently introduced an optical realization of next-generation reservoir computing (NGRC)[31] that bypasses the random-projection step entirely, replacing it with a deterministic nonlinear feature map. Çarpınlıoğlu and Teğin[30] explored the orthogonal direction of replacing the random scattering medium with a genetic-algorithm-optimized one, demonstrating that task-specific reservoirs outperform untuned ones on the same hardware. Each of these approaches is distinct from ours in its objective: the silicon-photonic work by the Huang group optimizes hardware integration, the Gigan-group optical NGRC modifies the reservoir computing framework itself, and the optical network by Çarpınlıoğlu and Teğin is task-specifically tuned. Our contribution is complementary — an explicit, transferable map of the design landscape of conventional, untrained, free-space scattering reservoirs. We expect the three design principles of Section 3.1 to inform each of these directions: the throughput-optimized silicon platform of Ref.[15] should still benefit from operating near $\lambda_{\max} \approx 0$; the NGRC framework of Ref. [31] still depends on input coupling and projection density; the task-tuned reservoirs of Ref. [30] still trade off connectivity. The relationship between conventional and trained-physical-layer architectures, including the deep diffractive networks of ref [28] and their successors[29], remains an open and complementary research question on which our design landscape is silent.

*3.5. Future Directions*

The natural extensions of the present work fall along four avenues. First, the optoelectronic prototype should be transposed to a fully-optical architecture in which the scatterer is placed inside a delayed optical cavity, eliminating the electronic round-trip and allowing reservoir-state evolution to occur at the natural timescale of optical propagation instead of the timescale technically constrained by modulation and detection speed. Such a transposition would place the design landscape mapped here directly into the bandwidth regime of ref [15] and beyond. Second, an attractive alternative engineering route in the optoelectronic workflow is to achieve higher processing throughput by replacing the liquid crystal SLM with a faster optical modulator[17], such as digital micromirror devices and MEMS-based phase light modulators. The three design principles articulated here transfer immediately, with the practical consequence of substantially higher state-update rate. Third, the reservoir explored here is dynamically homogeneous and topologically random; recent work on modular reservoir architectures realized in atomic-switch nanowire networks[11,36,37] suggests that introducing structured heterogeneity into the scatterer phase profile may unlock multitask capacity beyond that of a single-task-tuned reservoir. A complementary direction, motivated by the recent observation of bifurcation cascades in pulse-amplitude-modulated photonic reservoirs[38], is to introduce slow adaptive feedback into the scatterer such that $\lambda_{\mathrm{max}}$ tracks the changing statistics of the input — effectively converting the static design principle of Section 3.2 into a flexible operation control. We anticipate that the design landscape mapped here will provide the quantitative scaffold on which these extensions are evaluated.

## 4. Experimental Section

### *4.1. Optical Setup*

The experimental setup consists of a coherent light source, a phase-modulating display, a phase scatterer, a camera, and 4-f relay optics (Figure 1b). A coherent diode-pumped solid-state laser (wavelength 532 nm; Cobolt Samba, HÜBNER Photonics, Germany) serves as the light source. A reflective phase-only spatial light modulator (X10468, Hamamatsu Photonics K.K., Japan) projects the input-coupled reservoir state as a phase pattern on the optical field; a second SLM of identical specification operates as the reconfigurable scatterer that intermixes the field across reservoir nodes. A complementary metal–oxide–semiconductor (CMOS) camera (LT425, Teledyne Technologies, USA) detects the scattered intensity and provides the next reservoir state. Each reservoir node is mapped one-to-one between a 6×6 super-pixel of the input SLM and a single CMOS pixel; the mapping is calibrated experimentally by identifying the camera pixel that is most central to the region modulated by each super-pixel.

To suppress multi-pass reflections that compromise temporal stability, the modulated field is filtered at the aperture plane on top of a carrier spatial frequency, equivalent to adding a linear phase ramp to the desired phase profile. The binary aperture mask removes the direct-current term and other reflection components; the scatterer SLM is mounted in a moderately tilted configuration to further suppress retroreflection.

### *4.2. Reservoir Dynamics Control*

The reservoir's dynamical regime is controlled by a single physical parameter — the multiplicative gain to $\mathbf{W}_{\mathrm{rec}}$ in Equation (3), equivalent to the overall light intensity received per state update. Experimentally, the multiplicative coefficient is tuned by the camera exposure time $T$; in simulation, it is implemented as a numerical multiplier applied to the optical field intensity before the Clip step. Because shot noise was found to be insignificant across the operating range (verified by repeating the dynamical instability sweep at fixed $T$ with varied averaging), $T$ acts as a clean dynamical knob rather than a noise-amplitude parameter. Throughout this work, "exposure time" refers to the experimental control variable and "numerical multiplier" to its dimensionless simulation analogue.

### *4.3. Scatterer Phase Profile Generation*

The scatterer SLM displays a phase pattern that produces a uniform far-field intensity profile within a target diffusing angle $\theta$, generated by an iterative algorithm related to the Gerchberg–Saxton method[39]. The procedure is initialized in the Fourier domain with uniform amplitude and random phase within $\theta$. After backpropagation to the scatterer plane, the amplitude is replaced with unity (phase-only constraint). Forward propagation back to the Fourier plane is followed by amplitude replacement with unity inside $\theta$ and zero outside (angular constraint). The mean squared error between the resulting angular intensity profile and the diffusing-angle-bounded uniform target is evaluated each iteration; the pattern with the lowest MSE is retained (Figure S2).

*4.4. Input Encoding*

We use a tilde decoration to denote the input-coupling representation that makes Equation 3 dimensionally consistent: $\tilde{\mathbf{x}}(t) \in \mathbb{R}^M$ is the M-dimensional tiled input formed by cycling the L-dimensional $\mathbf{x}(t)$ through the reservoir nodes that the input couples to, and $\widetilde{\mathbf{W}}_{\text{in}} \in \mathbb{R}^{M\times M}$ is the corresponding diagonal input-coupling matrix. More specifically,

$$\tilde{x}_i(t) = \begin{cases} x_{(k-1) \bmod L}(t) & \text{if } \exists k \in \{1, \dots, \rho M\} \text{ such that } i = m_k \\ 0 & \text{otherwise} \end{cases}, \quad \text{(Eq. 8)}$$

and

$$\left(\widetilde{W}_{\text{in}}\right)_{ij} = \begin{cases} w & \text{if } i = j \text{ and } \exists k \in \{1, \dots, \rho M\} \text{ such that } i = m_k \\ 0 & \text{otherwise} \end{cases}, \quad \text{(Eq. 9)}$$

where $\{m_1, \dots, m_{\rho M}\}$ specifies the reservoir indices that are coupled with the input.

For tasks with scalar input — including memory-capacity characterization and Mackey–Glass prediction — all input-infused reservoir nodes receive an identical scalar. The input scalar $u_t$ is linearly rescaled to the interval $[0,1]$ across the entire sequence. The fraction of input-infused nodes (the $\mathbf{W}_{\text{in}}$ nonzero ratio $\rho$) and the magnitude $w$ of nonzero entries are the two parameters swept in Figure 3.

For spoken-digit recognition, each utterance is processed into a log-scaled mel spectrogram in three steps: short-time Fourier transform, mel-frequency filtering, and logarithmic rescaling (Figure 6a). The STFT uses a 30 ms Hamming window with a 15 ms hop; the magnitude squared yields the spectrogram. Mel-frequency filtering subsamples the spectrogram onto 32 perceptually equidistant pitch bands centered at

$$f_{\text{cen}}(k) = 700 \cdot \left(10^{\frac{k}{2595}} - 1\right), \quad \text{(Eq. 10)}$$

where each band is a triangular window peaking at the band center and vanishing at the adjacent pitch band centers. The log-mel signal is then channel-wise linearly rescaled so that each mel index spans $[0,1]$ across the dataset, mitigating the loss of moderate signal under bit-depth quantization.

*4.5. Maximal Lyapunov Exponent Measurement*

The maximal Lyapunov exponent $\lambda_{\max}$ characterizes the reservoir's dynamical instability. In contrast to the formal echo-state property, which concerns convergence under input drive, the autonomous (zero-input) maximal Lyapunov exponent measured here characterizes dynamical instability under perturbation and serves as an experimentally accessible proxy for the operating regime. We adopt the perturbation-propagation definition[40]:

$$\lambda_{\max} = \frac{1}{K}\sum_{t=1}^{K} \log \frac{\| \mathbf{h}'(t) - \mathbf{h}(t) \|}{\| \mathbf{h}'(t-1) - \mathbf{h}(t-1) \|}, \quad \text{(Eq. 11)}$$

where $\mathbf{h}'$ is the trajectory obtained from $\mathbf{h}(t=0)$ perturbed by a small displacement of $2^{-5}M^{1/2}$ at $t = 0$ and $\|\cdot\|$ is the Euclidean norm. Both trajectories evolve under zero input. To prevent perturbation saturation, the perturbation is renormalized to its initial magnitude after every update. Each $\lambda_{\max}$ value reported in Figure 2(a) is obtained by averaging over $K = 80$ time steps with 5 and 1 random initial perturbation directions for simulation and experiment, respectively.

*4.6. Memory Capacity Measurement*

The linear memory capacity quantifies how faithfully past inputs can be reconstructed from the current reservoir state[41]:

$$\mathrm{MC} = \sum_{\tau=1}^{\tau_{\max}} \frac{\mathrm{Cov}^2(x(t-\tau), \hat{y}_\tau(t))}{\mathrm{Var}(x(t))\,\mathrm{Var}(\hat{y}_\tau(t))}, \quad \text{(Eq. 12)}$$

where $y_\tau$ is the output element trained to reconstruct $u_{t-\tau}$, and the upper limit $\tau_{\max} = 2M$ exceeds the maximal observed memory across all parameter sweeps. The input sequence is independent and identically distributed uniform random in $[0,1]$ with length 1000 for training and 500 for testing.

*4.7. Readout Training*

Assuming $t = 1, 2, \ldots, t_{\mathrm{tr}}$ as the time indices over the training sequence, the reservoir states can be represented with the matrix $\mathbf{H} \in \mathbb{R}^{M \times t_{\mathrm{train}}}$, of which each column is the state $\boldsymbol{h}$. Similarly, the corresponding ground-truth outputs can be formed into the matrix $\mathbf{Y} \in \mathbb{R}^{N \times t_{\mathrm{train}}}$. Given $\mathbf{H}$ and $\mathbf{Y}$, the readout weight matrix is obtained by Moore–Penrose pseudoinverse:

$$\mathbf{W}_{\mathrm{out}} = \begin{cases} \mathbf{Y}\mathbf{H}^{\mathrm{T}}(\mathbf{H}\mathbf{H}^{\mathrm{T}} + \varepsilon \mathbf{I}_M)^{-1} & \text{if } t_{\mathrm{train}} \geq M \\ \mathbf{Y}(\mathbf{H}^{\mathrm{T}}\mathbf{H} + \varepsilon \mathbf{I}_T)^{-1}\mathbf{H}^{\mathrm{T}} & \text{if } M > t_{\mathrm{train}} \end{cases}. \quad \text{(Eq. 13)}$$

The classification readout in Figure 6 and Section 2.6 uses one-hot encoded $\mathbf{Y}$ of dimension $N = 10$, with predicted class assigned by averaging the predicted output over the duration of each utterance and selecting the maximal-response class.

*4.8. Numerical Simulation*

Numerical simulation reproduces the optical setup of Section 4.1 in the discrete domain. The complex field is sampled on a square super-pixel or pixel grid; SLM phase modulation is applied by element-wise complex exponentiation; the scattering is numerically simulated as a Matrix multiplication by the complex angular spectrum of the scatterer phase profile generated in Section 4.3; intensity detection is the squared magnitude with Clip applied at the experimental saturation threshold. Five independent random scatterer realizations are simulated for each parameter setting and averaged to obtain the values reported in Figures 2–4. Simulations were carried out in MATLAB, on a CPU of Xeon(R) E5-2630 (Intel(R), United States).

*4.9. Phase SLM Rationale*

We adopted phase-only spatial light modulators for both the input encoding and the scatterer for three reasons. First, phase modulation incurs no absorption loss, preserving the overall photon budget that controls the dynamical regime characterized in Section 2.2. Second, phase encoding contributes nontrivially to reservoir dynamics through the scatterer, generating richer state evolution than amplitude-only modulation alone. Third, the present study aims not to deliver an optimized reservoir-computing product but to map the design landscape of light-scattering RC; reconfigurable phase SLMs serve as a commonly deployable design knob. Digital-micromirror-device-based intensity modulation remains an attractive alternative for higher-throughput deployment and is compatible with the design principles established here.

*4.10. Demonstration parameters (cross-section reference)*

| Parameter | §2.2 (dynamics) | §2.3 (input coupling) | §2.4, §2.5, §2.6 (interconnectivity, demos) |
|---|---|---|---|
| $M$ | 400 (20×20) | 400 (20×20) | 10,000 (100×100) |

| Parameter | §2.2 (dynamics) | §2.3 (input coupling) | §2.4, §2.5, §2.6 (interconnectivity, demos) |
|---|---|---|---|
| $\theta$ | 0.2° | 0.2° | 0.2° |
| $\rho$ | 21% | swept | 21% |
| $w$ | 0.25 | swept | 0.25 |
| $T$ | swept | | tuned to $\lambda_{\max} \approx 0$ |

The same parameter conventions are used across Sections 2.2-2.6; the values that vary across sub-sections are summarized below.

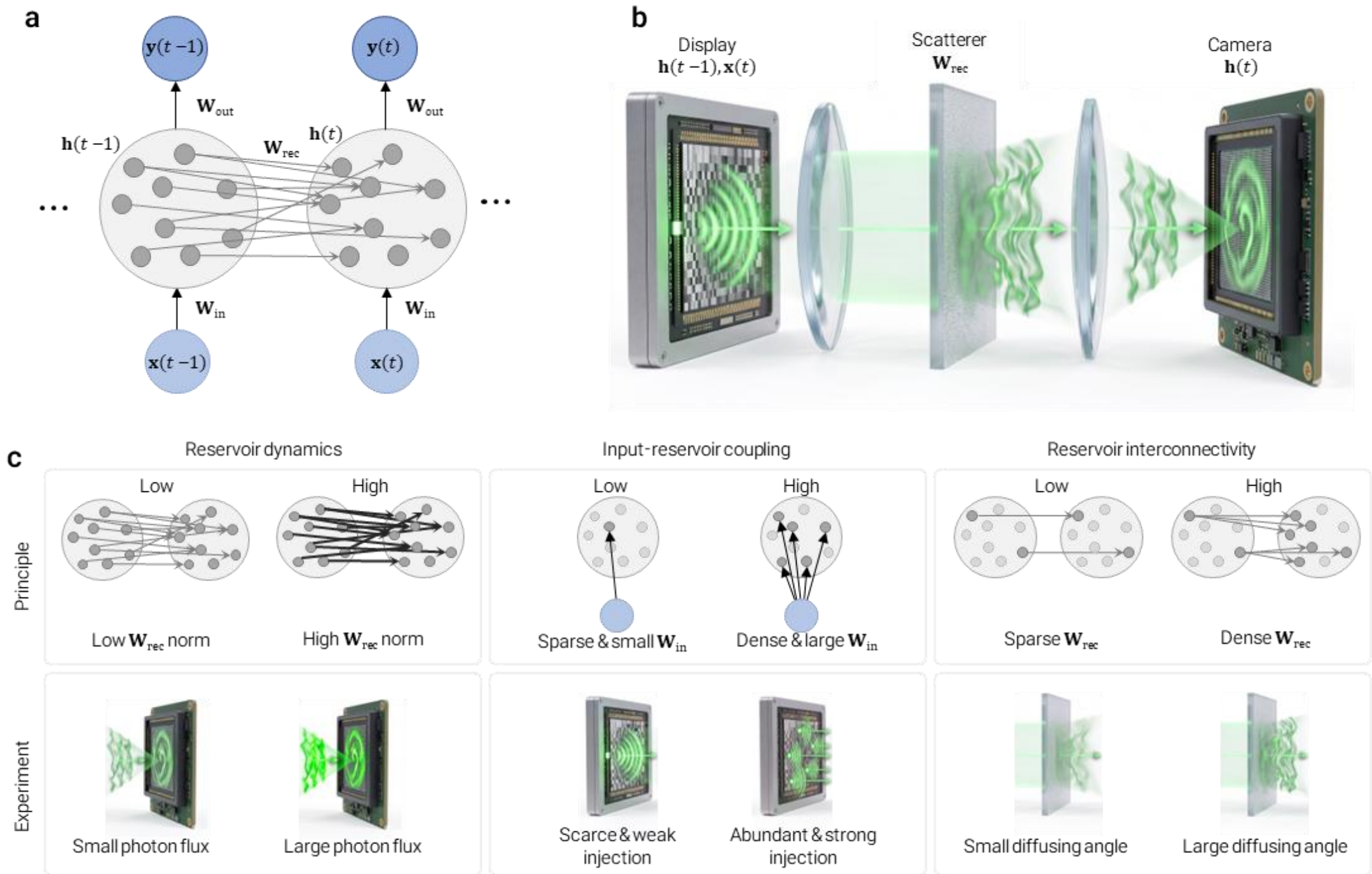

**Figure 1. Light-scattering reservoir computing platform and its three-axis design space.** (a) Conceptual illustration of a recurrent neural network (RNN) or reservoir computing system. At each time $t$, the reservoir state vector $\mathbf{h}(t)$ is obtained by the input vector $\mathbf{x}(t)$ and previous state $\mathbf{h}(t-1)$, which are rearranged to $\mathbf{h}(t)$ through $\mathbf{W}_{\text{in}}$ and $\mathbf{W}_{\text{rec}}$, respectively. (b) Simplified schematic of an optoelectronic reservoir computing system based on a 4-$f$ imaging system hardware. $\mathbf{x}(t)$ and $\mathbf{h}(t-1)$ are linearly fused with $\mathbf{W}_{\text{in}}$, encoded as a phase pattern on the display spatial light modulator (SLM), propagated through a reconfigurable scatterer $\mathbf{W}_{\text{rec}}$ realized by a second SLM, and read out by a CMOS camera as the next state. (c) Physical control axes mapped in this work: reservoir dynamics (related to $\|\mathbf{W}_{\text{rec}}\|$ controlled by camera exposure), input-reservoir coupling (controlled by adjusting effective number of $\mathbf{W}_{\text{in}}$ entries and magnitude), and reservoir interconnectivity (related to the degree of connection in $\mathbf{W}_{\text{rec}}$ controlled by the angular spectrum width of the scatterer.

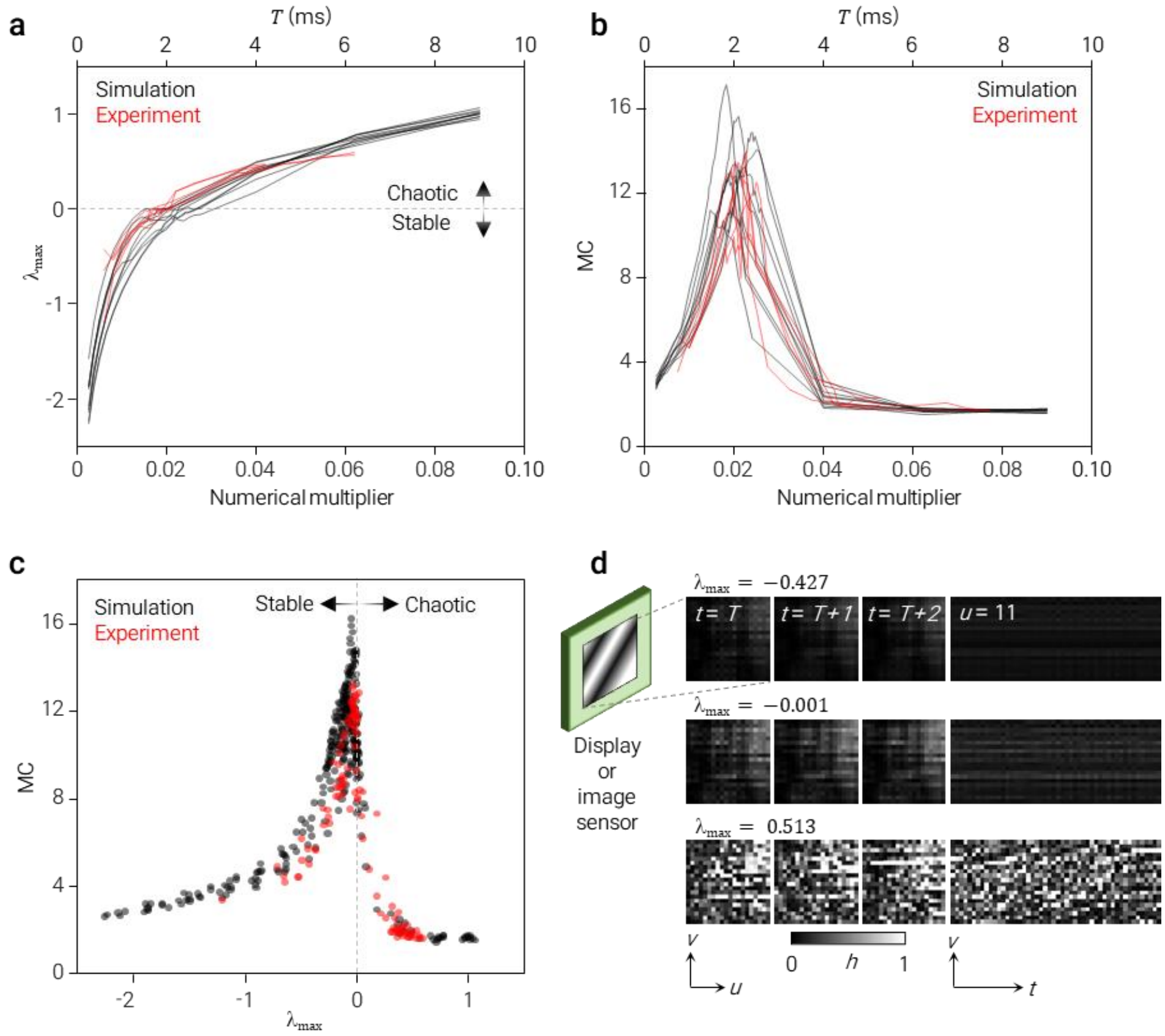

**Figure 2. Memory peaking at the dynamic criticality of the light scattering reservoir.** (a) Maximal Lyapunov exponent $\lambda_{\max}$ resulting from a range of camera exposure time, in numerical simulation (black) and experiment (red). Increasing exposure promotes the overall activation of the reservoir nodes, transitioning the reservoir from a stable regime ($\lambda_{\max} < 0$, converging attractor) through criticality ($\lambda_{\max} \approx 0$) into chaos ($\lambda_{\max} > 0$, exponential perturbation growth). (b) Memory capacity (MC) as a function of camera exposure. The simulation and experiment shows agreement in the behavior of $\lambda_{\max}$ throughout a range of camera exposure. (c) MC plotted versus $\lambda_{\max}$. Both simulation and experiment collapse onto a single sharp peak at $\lambda_{\max} \approx 0$; the dashed line marks the criticality, originally proposed for binary recurrent networks (Bertschinger and Natschläger, 2004) and recently confirmed experimentally in ion-gating (Nishioka et al., 2022) and spin-wave (Nishimura et al., 2024) reservoirs. (d) Simulated change of reservoir state for three dynamic regimes — stable (top), critical (center), chaotic (bottom). The stable reservoir approaches a certain state, the critical reservoir state varies within a range, and the chaotic reservoir state varies without a discernable restriction, illustrating the dynamical transition that underlies the MC peak.

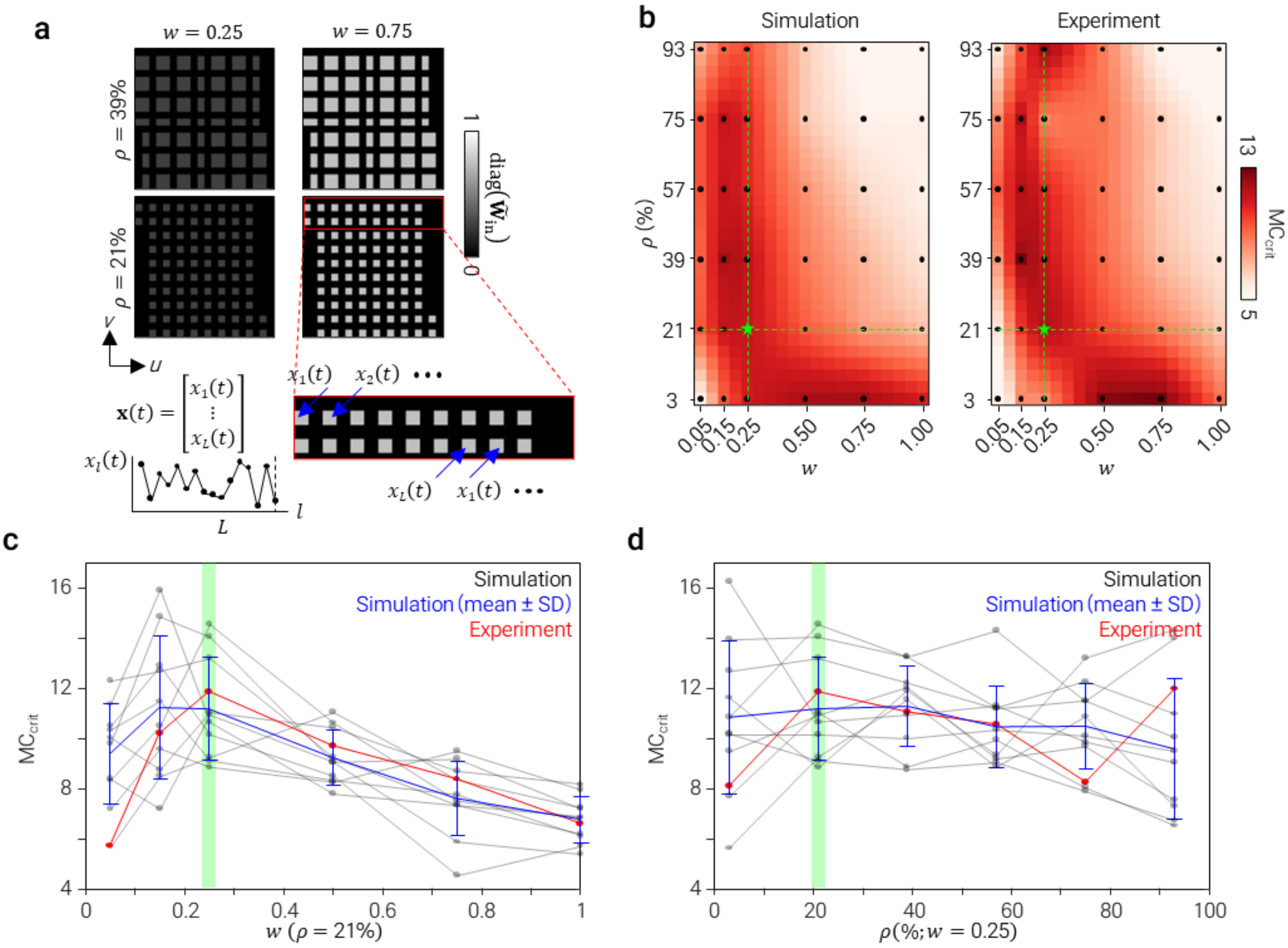

**Figure 3. Input–reservoir coupling and its memory-capacity optimum.** (a) Encoding scheme for the input vector $\mathbf{x}(t)$. Both the value and ratio of nonzero entries of $\mathbf{W}_{\text{in}}$ are controlled. An input of dimension $L$ is mapped onto the display super-pixels that correspond to the nonzero $\mathbf{W}_{\text{in}}$ entries repeatedly (blue arrows). (b) MC at the camera exposure closest to the dynamic criticality ($\text{MC}_{\text{crit}}$) as a function of the value ($w$-axis) and ratio of nonzero entries ($\rho$-axis) of $\mathbf{W}_{\text{in}}$. $\text{MC}_{\text{crit}}$ is suppressed in the lower-left corner (insufficient coupling leads to the recurrent dynamics silencing the input) and in the upper-right corner (excessive coupling leads to the input obscuring the previous reservoir state). The mapping is provided through natural-neighbor interpolation of investigated points (black dots). The central point along the range of high-$\text{MC}_{\text{crit}}$ parameters (green star) is chosen as the operating design used in subsequent demonstrations. (c,d) $\text{MC}_{\text{crit}}$ profiles along $w$-(c) and $\rho$-axes (d) of (b).

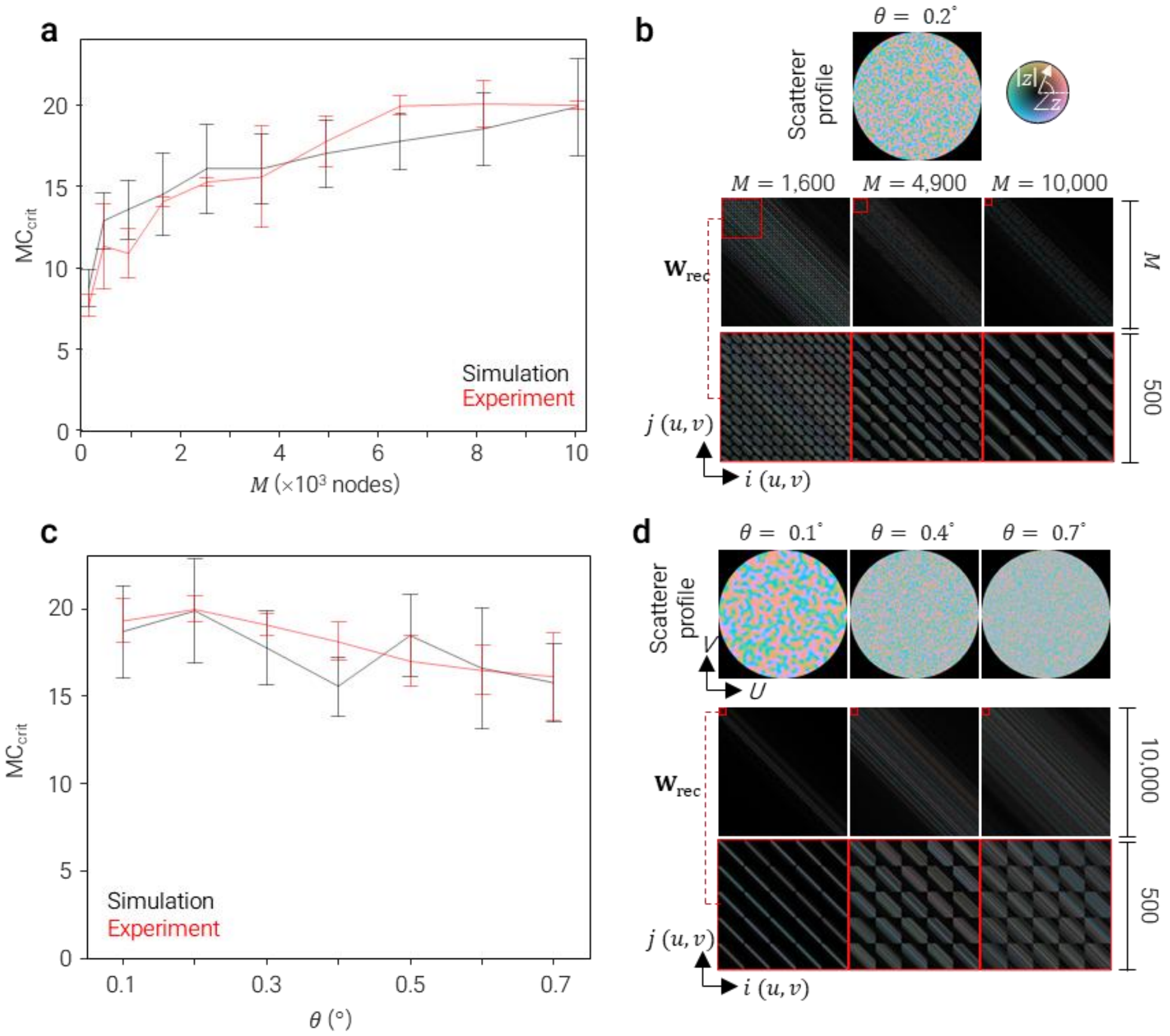

**Figure 4. Reservoir scale and scatterer interconnectivity exhibit a common saturating-design signature.** (a) $MC_{crit}$ as a function of reservoir size, defined as the number of effective display super-pixels and camera readout pixels, swept from $10 \times 10$ to $100 \times 100$. (b) Note that interconnectivity—visualized by the ratio of non-negligible entries of $\mathbf{W}_{\mathrm{rec}}$—decreases when using larger regions of the display or camera with an identical scatterer. (c) MC as a function of scatterer diffusing angle $\theta$, swept from 0.1° to 0.7°. The highest average $MC_{crit}$ appears at $\theta = 0.2°$ for both the experiment and simulation, indicating the optimal reservoir interconnectivity for memory. (d) Interconnectivity of $\mathbf{W}_{\mathrm{rec}}$ increasing with the scatterer diffusing angle, at a $100 \times 100$ reservoir.

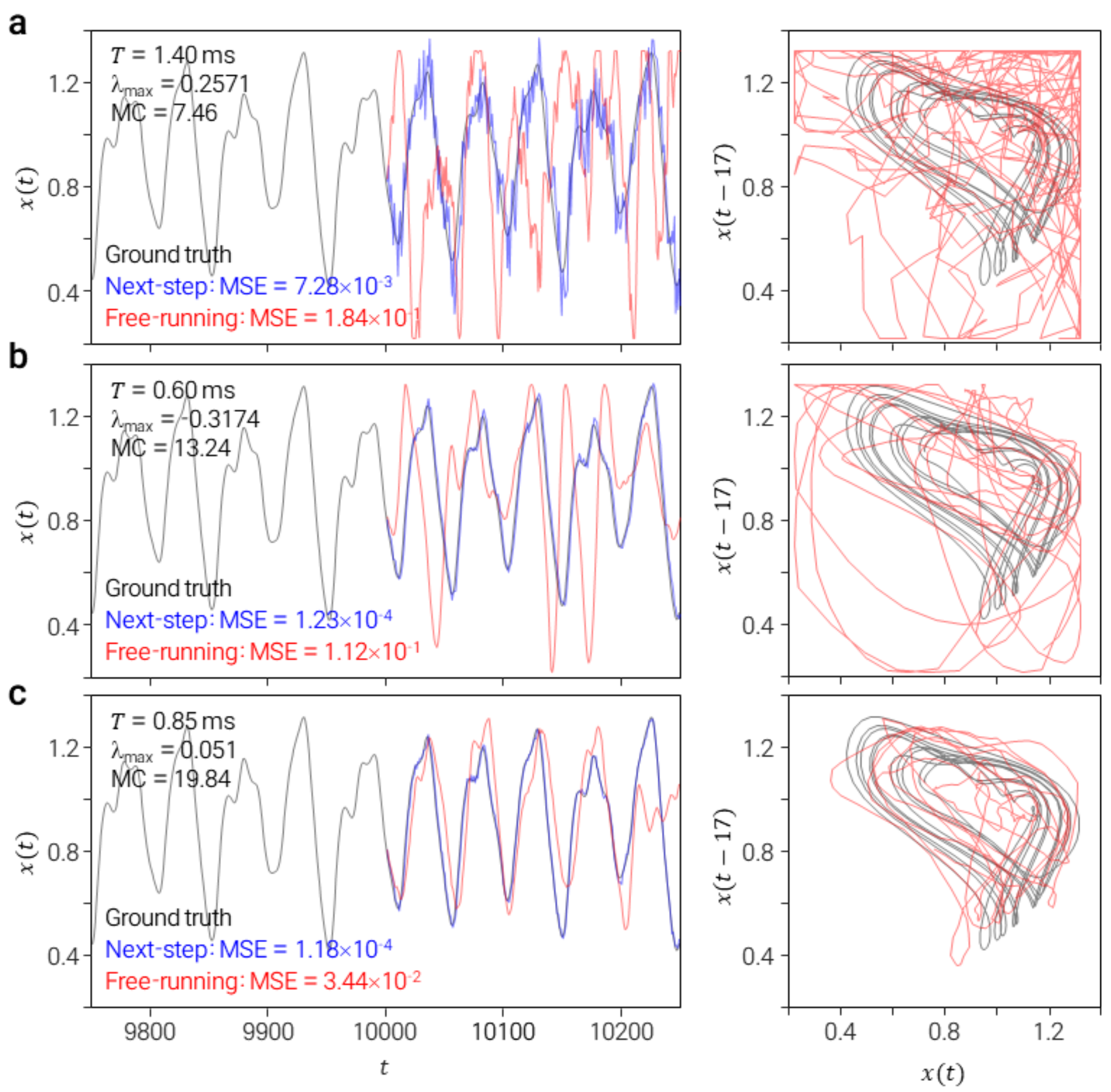

**Figure 5. Experimental chaotic time-series prediction across three reservoir-dynamics regimes.** The light scattering reservoir of size 10,000, configured at the input-reservoir coupling and interconnectivity identified in Figures 3 and 4, is operated at three detector exposure settings that place the dynamic (a) stable ($\lambda_{\max} < 0$), (b) critical ($\lambda_{\max} \approx 0$), and (c) chaotic ($\lambda_{\max} > 0$) regimes. For each regime, the left subpanel overlays the ground-truth Mackey-Glass time series (delay $\tau = 17$, gray) with the reservoir computing-based predictions (next-step prediction: blue, free-running prediction: red); the right subpanel shows the trajectory of ground truth and free-running prediction during the test time points ($t \in [10001, 10250]$) in the delay-embedded phase space $(z(t), z(t-\tau))$. While the stable reservoir is notably more accurate compared to the chaotic reservoir in next-step prediction, it fails to reproduce chaotic phase-space circulation of the original time-series. The reservoir at critical dynamics sustains recurrent phase-space circulation that resembles the ground truth, for over 200 time points.

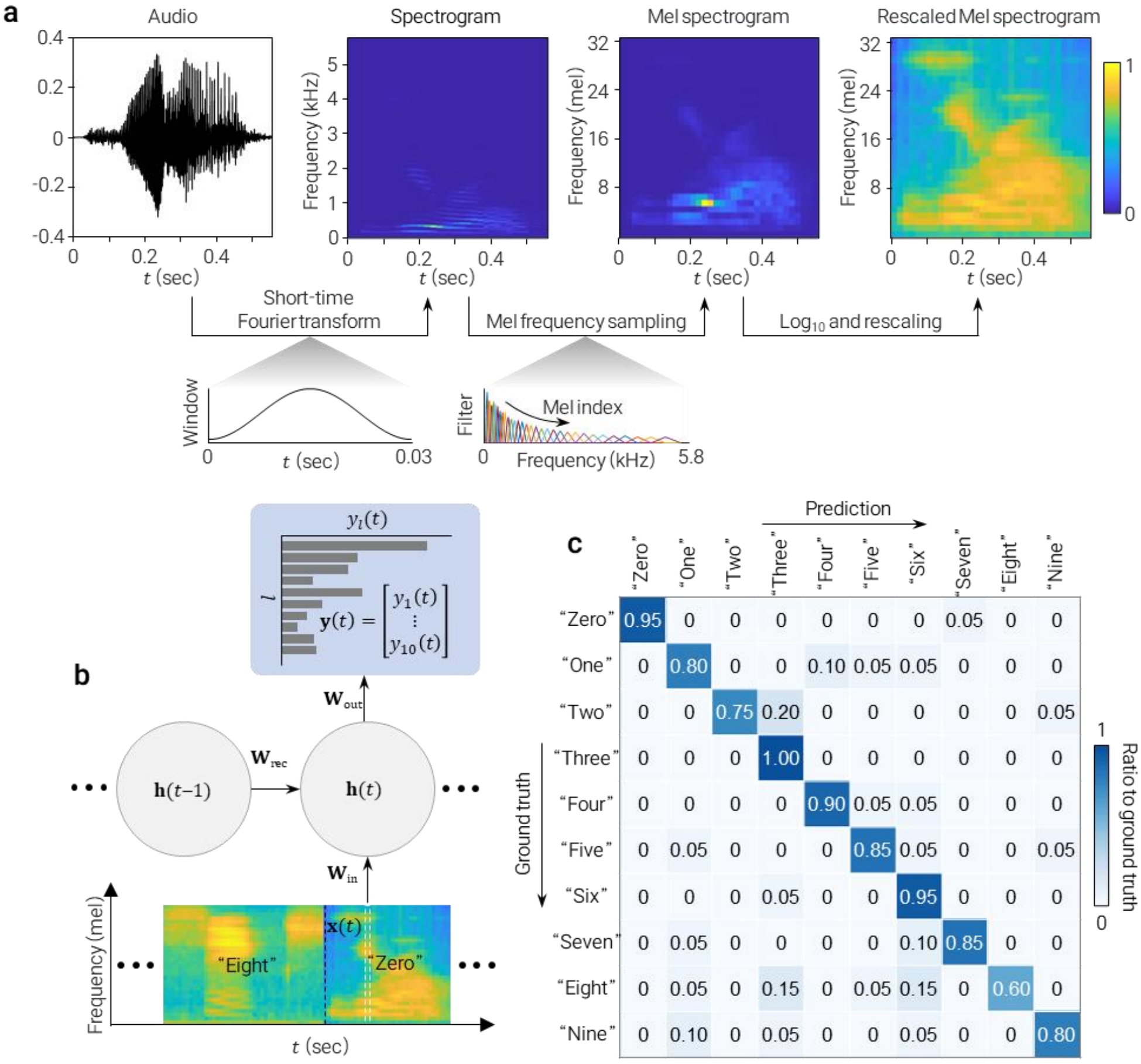

**Figure 6. Real-world spoken-digit recognition with the high-memory light scattering reservoir.** (a) Input data processing pipeline for a spoken word recording data. Each recording is preprocessed into a 32-channel log-mel spectrogram, which is streamed along time. (b) Schematic description of spoken-digit recognition task; the goal is to accurately indicate which digit is uttered, among the 10 digits. Each recording is classified based on the readout $\mathbf{y}(t)$ averaged over the duration of the recording. (c) Confusion matrix on the test set. Diagonal dominance indicates correct classification overall; the residual off-diagonal elements indicate digit pairs misrecognized by the reservoir (e.g., "two" misrecognized as "three", followed by "eight" misrecognized as "three" or "six"). The classification accuracy is 84.5% within the entire test set.

**Acknowledgements**
This work was supported by the National Research Foundation of Korea grant funded by the Korea government (MSIT) (RS-2024-00442348, RS-2022-NR068141), Korea Institute for Advancement of Technology (P0028463), and the Samsung Research Funding Center of Samsung Electronics under Grant (SRFC-IT1401-08). The authors thank KyeoReh Lee and Seungwoo Shin for fruitful technical discussions regarding the hardware realization.

**Conflict of Interest**
The authors declare no competing interests.

**Author Contributions**
G.K. conceptualization, methodology, software, formal analysis, investigation, data curation, visualization, writing — original draft. Y.K.P. conceptualization, supervision, project administration, funding acquisition, writing — review & editing.

**Data Availability Statement**
The data that support the findings of this study are available from the corresponding author upon reasonable request.

**Ethics, Patient Consent, Clinical Trial Registration**
Not applicable. This study does not involve human or animal subjects.

**Permission to Reproduce Material**
All figures are original to this work. No permission to reproduce third-party material is required.